# NEURO FUZZY MODELLING FOR PREDICTION OF CONSUMER PRICE INDEX

Godwin Ambukege, Godfrey Justo and Joseph Mushi

Department of
Computer Science and Engineering, University of Dar es Salaam, Tanzania.

## ABSTRACT

*Economic indicators such as Consumer Price Index (CPI) have frequently used in predicting future economic wealth for financial policy makers of respective country. Most central banks, on guidelines of research studies, have recently adopted an inflation targeting monetary policy regime, which accounts for high requirement for effective prediction model of consumer price index. However, prediction accuracy by numerous studies is still low, which raises a need for improvement. This manuscript presents findings of study that use neuro fuzzy technique to design a machine-learning model that train and test data to predict a univariate time series CPI. The study establishes a matrix of monthly CPI data from secondary data source of Tanzania National Bureau of Statistics from January 2000 to December 2015 as case study and thereafter conducted simulation experiments on MATLAB whereby ninety five percent (95%) of data used to train the model and five percent (5%) for testing. Furthermore, the study use root mean square error (RMSE) and mean absolute percentage error (MAPE) as error metrics for model evaluation. The results show that the neuro fuzzy model have an architecture of 5:74:1 with Gaussian membership functions (2, 2, 2, 2, 2), provides RMSE of 0.44886 and MAPE 0.23384, which is far better compared to existing research studies.*

## KEYWORDS

## 1. INTRODUCTION

Economists and policy makers around the globe put efforts on understanding economic factors, such as conditions in foreign trade, marketable surplus of agriculture, and noneconomic, to improve and maintain economic status of their respective countries [1]. Accurate prediction of macro-economic indicators have proven to support good understanding of economy. To date, various economic indicators of interest can predicted for specific sectors of economy include Consumer Price Index (CPI), inflation rates, Gross Domestic Product (GDP), birth rates, unemployment rates, and stock markets [2]. Among these economic indicators, CPI is the key economic indicator that measures the change over time in the purchasing cost of a fixed basket of goods and services that are consumed by a representative sample of households in a given country [3].

CPI is the measure of price level of goods and services in market basket that usually used to determine inflation rate in respective community. Various researches have proven that the use of CPI is helpful in predicting future developments in business cycle [4]. Thus, accurate prediction of CPI plays a significant role for better planning of economic strategies in almost every institution such as government, financial institutions, academia, investors in industries and agriculture, savings and credit cooperative unions, and individual consumer as well.

Various CPI prediction models have been proposed, however, their prediction accuracy is still low. This can be justified from the study by [5], [6], and [7]. The low prediction accuracy for these models can be caused by factors such as small volume of data used, improper selection of optimization techniques, the use of single prediction technique independently, selection for number of input variables or due to the expertise of the model designer.

Therefore, in this research study, the Artificial Intelligence (AI) approach was applied by combining neural networks and fuzzy logic to develop a neuro fuzzy model so as to improve prediction accuracy of CPI. The sample data was taken from Tanzania National Bureau of Statistics (TNBS).

## 1.1 CPI in Tanzania

Tanzania started to compile National Consumer Price Index (NCPI) on yearly basis since 1965, then on quarterly basis from 1974 to 1994 and from then to date the NCPI is calculated on monthly basis and released to the public on $8^{th}$ day of the subsequent month [8]. NCPI is also known as total CPI or all item index, for simplicity, in this paper, the CPI will mean NCPI. CPI is used to compute an average measure of price inflation for the household sector as a whole, to adjust wages as well as social security and other benefits to compensate for the changes, normally rising of consumer prices, it acts as one of macro-economic indicators in adjustment of government fees and charges, adjustment of payments in commercial contracts, and international comparison [9].

The CPI is the weighted average of the price of goods and services computed by taking the cost of market basket in a given year dividing the cost of market basket in a base year or reference year and the results multiplying by 100. For example if we have two goods A and B in the market basket and that CY denotes the expenditure of A and B in current year and BY denotes the expenditure of A and B in base year then, the CPI for current year is given by equation 1

CPI = (CY / BY) * 100                                         (1)

The CPI for each group is calculated using formula in equation 1, and the total CPI is then computed as the sum of the product of group CPI with its respective weight. If CPIc represents current CPI for one group and Wc denotes its associated weight, then, the total CPI (NCPI) for groups of goods and services *n* to *k* is calculated using equation 2.

$$NCPI = \sum_{n=1}^{k} CPIcWc$$          (2)

## 1.2 Adaptive Neural Fuzzy Inference Systems

A neural fuzzy model is a hybrid model which is found by combining Artificial Neural Networks and Fuzzy Logic techniques [10]. The neural fuzzy systems, also known as network based fuzzy systems, can be trained to develop IF-THEN fuzzy rules and determine membership functions for input and output variables of the system. Basically the neuro fuzzy system is a fuzzy system which contains membership functions that are tuned using training ability of neural networks. Research studies have shown that hybrid models perform better when solving a particular problem compared to an individual model alone [11], [12]. The integration of two techniques overcomes the constraints of individual model by hybridization of various methods. The motivation on developing a network based fuzzy system comes from the fact that, artificial neural networks are good at recognizing patterns, they have a clear established architecture with learning and generalisation ability. However, they are not good at explaining how they reach their decision. Neuro fuzzy models can be applied to solve problems in fields of medicine, economic forecasting, defence techniques by military and in engineering design.

## 1.3 Architecture of Neuro Fuzzy Systems

Network based fuzzy systems combine the merits of connectionist neural networks and fuzzy approaches as a soft computing component, and rule generation from ANN has become popular due to its capability of providing some insight to the user about the knowledge embedded within the network [13]. Hybrid models are important when considering wide range of application domains. The architecture of neuro fuzzy system is similar to multi-layered neural network with additional of two adaptive layers. For simplicity we assume neuro fuzzy system with two inputs ($x$ and $y$) and one output ($f$), the architecture consists of five layers as shown in Figure 1. Two IF THEN rules can be considered to represent this architecture.

Rule (1) IF x is $A_1$ and y is $B_1$ THEN $f_1 = p_1x + q_1y + r_1$.
Rule (2) IF x is $A_2$ and y is $B_2$ THEN $f_2 = p_2x + q_2y + r_2$.

Where: x and y are inputs, Ai and Bi are the fuzzy sets, fi are outputs, pi, qi, and ri are consequent parameters that are determined during training process.
The first layer accepts crisp inputs and converts to the corresponding membership functions (MF). This is an adaptive layer which fuzzifies inputs to membership grades of linguistic variables. The output $O_1$ of this layer is the fuzzy set of the inputs and is given by equation 3.

O1,i = $\mu_A i(x)$ for i = 1, 2
O1,i = $\mu_B i(y)$ for i = 1, 2         (3)

Where $\mu_A(x)$ and $\mu_B(y)$ are membership functions which can be of any type such as Gaussian membership function or Bell shaped. For example, for Gaussian MF, it is given by equation 1.2.

$$\mu A_i(x) = exp\left[\frac{-\,(C_i - x)^2}{2\sigma i^2}\right]$$     (4)

Where $ci$ and $\sigma i$ are centre and width of the $i$th fuzzy set $A_i$ respectively.

The second layer (represented by $\pi$) is the fuzzy rule layer and each neuron corresponds to a single fuzzy rule. A fuzzy rule neuron receives inputs from the fuzzification neurons that represent fuzzy sets in the rule antecedents. The layer calculates the weight of each rule. The output of this layer is calculated as the product of the incoming signal.

O2,i = wi = $\mu_{A,i}(x)$. $\mu_{Bi}(y)$ , i = 1, 2     (5)

The third layer (represented by N) is called normalized layer. This layer calculates the firing strengths of each rule from previous layers. The output of this layer is computed using equation 6

$$O3,i = w_i = w_i/(w_1+w_2) \text{ , } i = 1, 2 \tag{6}$$

The fourth layer is the output membership layer (y). Neurons in this layer represent fuzzy sets used in the consequent of fuzzy rules. The output of this layer is the product of the normalised firing strength of the rule obtained in previous layer and first order polynomial.

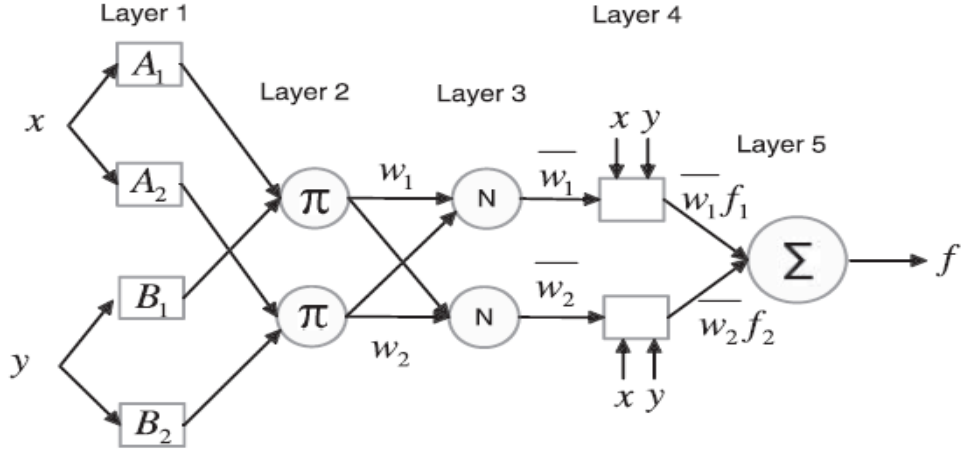

Figure 1: The Architecture of Neuro Fuzzy System

$$O4,i = w_if_i = w_i(p_iv + q_id + r_i) \text{ , } i = 1, 2 \tag{7}$$

Where $f_i$ is the node function, $p, q, r$ are consequent parameters which are determined during training process.

The last layer is the defuzzification layer which represents a crisp output of the neuro fuzzy system. This layer sums up all incoming signal and is computed using equation 8.

$$O5,i = \sum w_if_i = \sum w_if_i/(w_1+w_2), i =1, 2 \tag{8}$$

## 2. METHODOLOGY

### 2.1 Data Preparation

The first step was to prepare the data. Sample data were CPI data points for sixteen years from January 2000 to December 2015. The data collected were transformed into matrix vector having one row and many columns. The data were pre-processed so as to be acceptable by the Matrix Laboratory (MATLAB) computing platform. The data were separated into training samples (95%) and testing samples (5%). The data from January 2000 to March 2015 were taken as training data sets and from April 2015 to December 2015 were for model validation. One month ahead forecast technique was selected for both in sample and out of sample data. For instance, CPI data from January 2000 to May 2000 were used to forecast CPI for June 2000, from February 2000 to June 2000 was used to forecast CPI for July 2000 and so on. This means that, the forecasted data point at one month were taken as independent variable in predicting CPI for the next month. In this research work, MATLAB software was used for simulating the experiments in the process of designing and testing neuro fuzzy model. Root Mean Square Error (RMSE) and Mean Absolute Percentage Error (MAPE) were used as error metric as proposed by [14], [15], [16].

## 2.2 Training and Testing the Model

The process of model development can be explained using Figure 2. After defining the inputs and output sets for both training and testing, the parameters for membership functions (MF) were initialized, the appropriate number of input variables and output variable were selected. Both commands (genfis1 and genfis2) were used for initializing membership function parameters to generate Fuzzy Inference System (FIS) and the performance for each method was recorded.

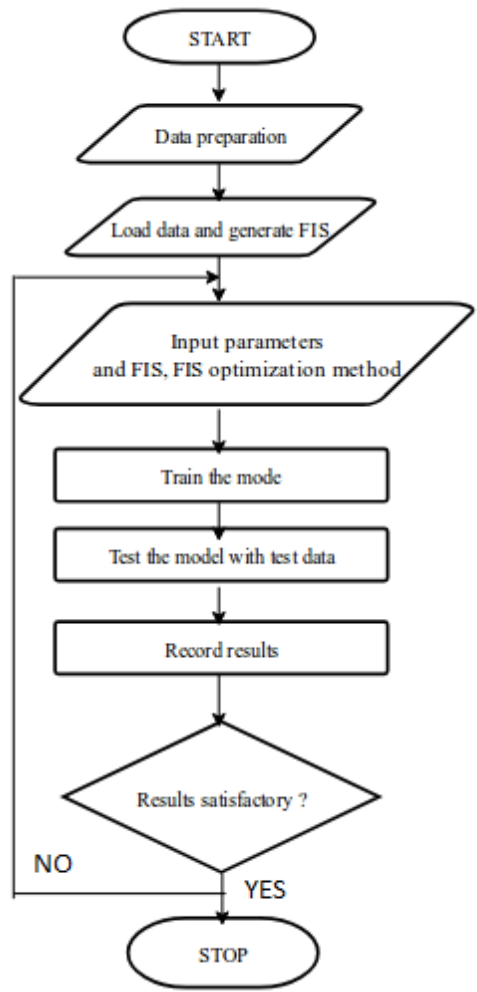

Figure 2: Model development process

Both Back propagation (BP) and hybrid method which combines BP and least square were used as training optimization methods. The training process was done using anfis command and different number of training epochs were tested while observing the values of Root Mean Square Error (RMSE) and Mean Absolute Percentage Error (MAPE). The error goal was set to zero, initial step size to 0.01, step decrease rate 0.9 and step size increase rate 1.1. These training parameters are important in controlling the learning process of the model.

After finishing the training process, the final membership functions and training error from the training data sets were produced. The command evalfis was applied to study the performance by applying inputs data to the fuzzy system without output data. The output produced here is the system output and was compared against the desired output for training and test data sets. The training errors were calculated using the formula in equation 9 and 10.

$$RMSE = \frac{\sqrt{\sum_{i=1}^{n}(yi - xi)^2}}{N}$$

(9)

Where $xi$ and $yi$ are the values of the $ith$ observations in x (predicted value) and y (target value), respectively and N is the total number of observations.

$$MAPE = \frac{1}{n}\sum_{t=1}^{n}Abs\left(\frac{At - Ft}{At}\right) * 100$$

(10)

Where Abs denotes absolute value, $A_t$ is the actual value (target) and $F_t$ is the forecasted value, n is the number of observations.

If the training error is not satisfactory the training process repeats. A number of simulation experiments were conducted while changing the training parameters such as number of inputs, type of membership functions, membership grades, and number of training epochs, in each experiment so as to obtain satisfactory training results. Table 1 shows sample training parameters.

Table 1: Neuro Fuzzy Training Parameters

| Sn | Parameter | Test cases | | | |
|---|---|---|---|---|---|
| 1 | Fuzzy system generation method | genfis1 | genfis2 | | |
| 2 | Type of Membership Function (MF) | Triangular | Trapezoidal | Generalized Bell | Gaussian |
| 3 | Number of inputs-outputs | 2-1 => 6-1 | | | |
| 4 | Number of MF per input | 2 => 4 | | | |
| 5 | Training algorithm | Backpropagation | Hybrid (Backpropagation + Least Square | | |
| 6 | Number of training epoch | 30 => 5,000 | | | |

# 3. RESULTS AND DISCUSSION

## 3.1 Training and Test Results

Table 2 shows the results of sample simulation experiments conducted to train and test different neuro fuzzy models using the MATLAB software. This table does not include all simulation experiments that were conducted; the table indicates few of them.

Table 2: Sample results for simulation experiments

| Sn | MF | MFs per input | Epochs | Architecture | RMSE Training | RMSE Testing | MAPE Testing |
|----|----|----|----|----|----|----|----|
| 1 | Gaussian | 3,3,3,2 | 500 | 4:119:1 | 0.36528 | 15.4779 | 8.312542 |
| 2 | Triangular | 3,3,3,3 | 450 | 4:174:1 | 0.44014 | 112.042 | 64.6368 |
| 3 | Triangular | 3,2,2,2,2 | 1000 | 5:107:1 | 0.40803 | 54.0926 | 31.12558 |
| 4 | Generalized bell | 3,3,2,3,2 | 750 | 5:229:1 | 0.18846 | 60.3491 | 34.97454 |
| 5 | Generalized bell | 3,3,2,2,2 | 1500 | 5:156:1 | 0.19256 | 65.0363 | 34.08347 |
| 6 | Triangular | 2,2,2,2 | 1000 | 4:40:1 | 0.59783 | 168.6596 | 97.1052 |
| 7 | Generalized bell | 2,3,3,2,2 | 500 | 5:156:1 | 0.24493 | 45.7328 | 26.28622 |
| 8 | Gaussian | 2,2,2,2,2 | 1000 | 5:74:1 | 0.40345 | 0.68746 | 0.332627 |
| 9 | Gaussian | 2,2,2,2,2 | 650 | 5.74:1 | 0.64241 | 0.44886 | 0.233839 |
| 10 | Generalized bell | 2,2,2,2,2 | 650 | 5:74:1 | 0.41263 | 1.183 | 0.649738 |
| 11 | Generalized bell | 2,2,3,3 | 650 | 4:82:1 | 0.45368 | 20.5294 | 12.25982 |

On average, genfis1 produced better results than genfis2. The triangular membership functions produced a higher forecasting error compared to other membership functions, Gaussian membership functions (2, 2, 2, 2, 2) in record number 9, was selected as the best model since it performed well compared to others. The model has RMSE of 0.44886 and MAPE of 0.233839. The bell shaped having 5 inputs and membership functions (3,3,2,3,2) shown in record number 4 produced good training results at 750 epochs but was not selected due to huge forecasting error (RMSE of 60.3491). Over fitting was noticed when the number of training epochs were increased to 900 and 1000 for Gaussian membership functions (2,2,2,2,2) of which the results was not good during testing (RMSE of 0.68746 and MAPE of 0.332627).

The obtained results imply that, some membership functions such as triangular do not suit well in modeling some applications such as those related to predictions, for this case CPI prediction. Also when modeling neural fuzzy, the model with low error during training does indicate it is the best model, for example, in this case; the training error was 0.64241 in record 9 of Table 4.2 and 0.18846 in record 4. The model in record 4 was not selected as the best model due to high error during testing (60.3491).

In this work, a new Sugeno type neuro fuzzy model was developed to predict the consumer price index. The architecture of the proposed model is shown in Figure 2. The selected Sugeno FIS has five layers, having 5 neurons in input layer, 10 neurons in input membership functions, 32 IF THEN rules, 32 neurons in output membership functions and 1 neuron in output layer. This architecture is represented as 5:74:1 in three layers model which means 5 input neurons, 74 neurons in hidden layers and 1 output neuron. The architecture has 2 Gaussian membership functions per input, which correspond to the set of (low and high) for CPI. Figure 3 shows portions of the rule viewer (rules) to represent the relationship between inputs and outputs. This indicates that, for any given five consecutive months' CPI, with the proposed model, it is possible to predict the next month CPI.

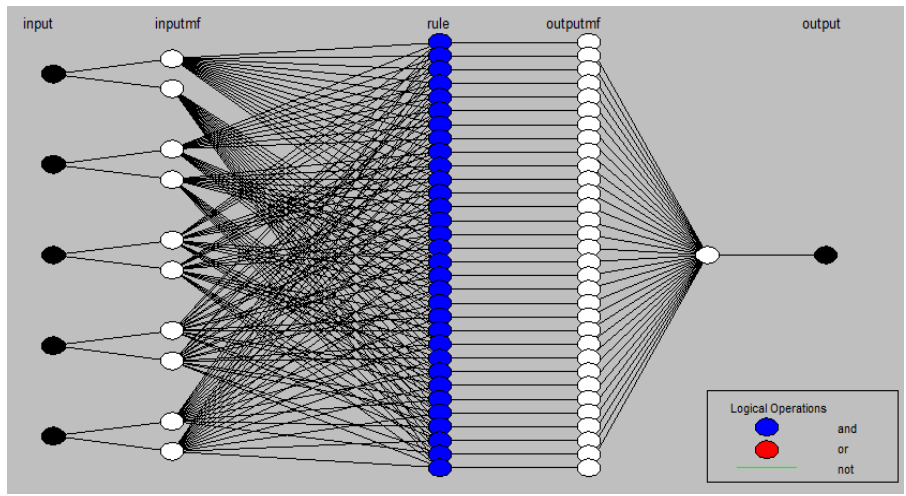

Figure 2: Architecture of proposed model

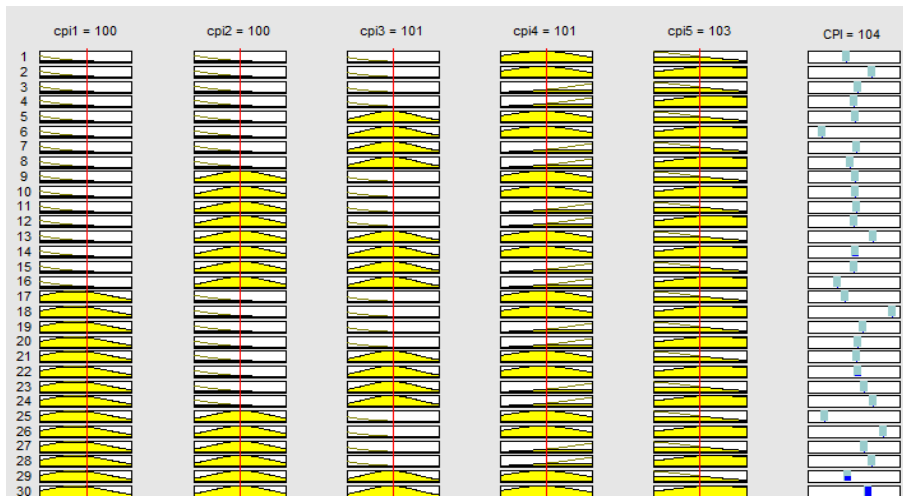

Figure 3: Rule viewer to interpret the rules

## 3.2 Prediction Results by Proposed Model

Table 3: Forecasting Results for Selected Neuro Fuzzy Model

| Month | Target | Predicted | Error |
|-------|--------|-----------|-------|
| Apr-15 | 157.21 | 156.5460181 | 0.663981936 |
| May-15 | 157.86 | 157.8974576 | -0.037457551 |
| Jun-15 | 158.12 | 158.2561941 | -0.136194111 |
| Jul-15 | 158.78 | 158.5041795 | 0.275820524 |
| Aug-15 | 158.81 | 159.3802297 | -0.570229734 |
| Sep-15 | 159.04 | 159.2140768 | -0.174076814 |
| Oct-15 | 159.17 | 159.5692438 | -0.399243834 |
| Nov-15 | 160.49 | 159.6562615 | 0.833738462 |
| Dec-15 | 161.24 | 161.4976339 | -0.257633946 |

Table 4 Performance of Chosen Neuro Fuzzy Model

| Model | Membership functions | RMSE | MAPE |
|---|---|---|---|
| Neuro fuzzy (5:74:1) | Gaussian (2, 2, 2, 2, 2) | 0.44886 | 0.23383928 |

As shown in Table 3, the chosen model was then used to predict CPI for 9 months from April 2017 to December 2015. As shown, the prediction errors are small. Table 4 shows the average prediction errors for 9 months in RMSE (0.44886) and MAPE (0.233839)

### 3.3 Results Discussion

In this research work, the neuro fuzzy models developed by [7], [17] and [18], have used as the benchmark. The results of the developed neuro fuzzy model was compared with these research works on CPI prediction and found that the proposed model performs better.

[7] researched on neuro fuzzy modelling technique using CPI data for a period of 15 years from January 2000 to January 2014. They used seven input variables and one output to forecast the US future CPI using subtractive clustering (genfis2) method for generating FIS. They trained the network for 10,000 epochs and obtained the values for RMSE = 0.837. The use of large number of inputs (7) per training pair, and genfis2 FIS initialization method during modelling process could be the cause of low prediction results.

In this research work has found that, in modelling neuro fuzzy for CPI prediction, Gaussian membership functions provide better prediction results when we use grid partition (genfis1) as FIS generation method, low number of inputs per training vector (5 in this case) and small number of membership functions per input (2 in this research). Table 5 shows prediction accuracy comparison for the proposed work with existing one.

Table 5 Comparison of Proposed Model with Other Models

| Prediction technique | RMSE | MAPE | Reference |
|---|---|---|---|
| Neuro fuzzy | 3.1865 | 2.9771 | (Ucenic and Atsalakis, 2009) |
| Neuro fuzzy | 1.4679 | | (Rosadi, Subanar, and Suhartono, 2013) |
| Neuro fuzzy | 0.829 | | (Enke and Mehdiyev, 2014) |
| Neuro fuzzy | 0.44886 | 0.23384 | Proposed |

# 4. CONCLUSION

This research study applied the concepts of artificial intelligence, specifically, the neuro fuzzy technique. The study has presented step by step procedures on how to develop the forecasting model for consumer price index using this technique. The data for this research work were taken from the Tanzania National Bureau of Statistics for 16 years from January 2000 to December 2015. The Root Mean Square Error and Mean Absolute Percentage Error were used as accuracy metrics to evaluate the performance of the developed models. The findings from this study were compared with existing research works on modelling neuro fuzzy for CPI prediction and showed that the proposed model provides better prediction accuracy.

This research work has shown an improvement on prediction accuracy of the consumer price index. The findings of this research should be used as the basis for building an application that can be used for predictions of consumer price index in real time environment. The appropriate authorities such as central banks and statistical bureaus which compile macro-economic indicators, specifically the consumer price index, should use the findings of this study as a guide in the prediction of consumer price index.

## AUTHORS

Godwin Ambukege received MSC (Computer Science) at University of Dar es Salaam (UDSM) in Tanzania, PGD (IT) at Amity University (India). He currently works as Senior Systems administrator at Teofilo Kisanji University in Mbeya Tanzania. He is also a member of Embedded and Intelligent systems research group at UDSM. His research interests include; Artificial intelligence, machine learning, expert systems and data analysis.

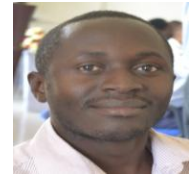

Dr. Godfrey Justo received his PHD (London), MSC (Math) Zimbabwe. He is Engagement and Management Advisor at Tanzania Data Lab (dLab). He is also a senior lecturer and expert in Software Engineering at UDSM, College of ICT. He has been working with various organisations such as World Bank, Japan International Cooperation Agency (JICA) and German Academic Exchange Service (DAAD) as consultant.

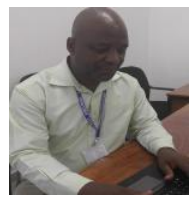

Joseph Mushi received his Msc. Degree in Communications and Information Systems Engineering at Wuhan University of Technology, Wuhan, P.R. China and Bsc. Computer Science at UDSM, Tanzania. He is the Facility Manager of Tanzania Data Lab (dLab) with background in Data and Statistical Analysis. He has been involved in numerous open data related activities as technical advisor, including capacity building programmes funded by World Bank and HDIF. He also holds an assistant lecturer position at the University of Dar es Salaam (UDSM), Department of Computer Science and Engineering.

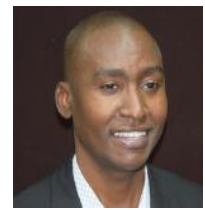